\documentclass[prb,twocolumn,showpacs,floatfix,superscriptaddress,amsfonts]{revtex4}
\usepackage{amsmath,amssymb,graphicx,bm,epsfig}
\usepackage{epstopdf,color}

\begin{document}
	
\title{Spin-fermion model for skyrmions in MnGe derived from strong correlations}

\author{Hongchul Choi}
\affiliation{Theoretical Division, Los Alamos National Laboratory, Los Alamos, New Mexico 87545, USA}

\author{Yuan-Yen Tai}
\affiliation{Theoretical Division, Los Alamos National Laboratory, Los Alamos, New Mexico 87545, USA}

\author{Jian-Xin Zhu}
\email{jxzhu@lanl.gov}
\affiliation{Theoretical Division, Los Alamos National Laboratory, Los Alamos, New Mexico 87545, USA}
\affiliation{Center for Integrated Nanotechnologies, Los Alamos National Laboratory, Los Alamos, New Mexico 87545, USA}

\begin{abstract}
MnGe has been reported as a candidate of three-dimensional (3D) skyrmion crystal in comparison to the two-dimensional (2D) skyrmion observed in most other B20 compounds like MnSi. In addition, the small-sized skyrmions in MnGe are desired properties for information storage. By performing the density functional theory (DFT) calculations and model simulations based on the DFT-informed tight-binding Hamiltonian, we explore the nature of the 3D skyrmion in MnGe.  By invoking a dual nature of $d$-electrons on Mn atoms, we propose a strong-correlation derived spin-fermion model with an antiferromagnetic coupling between the localized and itinerant moments. This model could explain the drastic difference of magnetic moments  between  MnGe and MnSi compounds.  In addition, we find that the 3D or 2D nature of skyrmions are dependent on the coupling strength.
\end{abstract}
	
\pacs{12.39.Dc,71.10.-w,71.15.Nc}
\maketitle
	

\section{introduction}
Mathematically, a skyrmion is a topological soliton solution known to occur in a non-linear field theory of hadrons in nuclear physics, originally proposed by Skyrme.~\cite{Skyrme:1962}  Nowadays the skyrmions are found to be relevant in condensed matter systems including quantum Hall systems,~\cite{SLSondhi:1993} liquid crystals,~\cite{DCWright:1989} and Bose condensates.~\cite{TLHo:1998} A  magnetic skyrmion  makes up a topological configuration of non-coplanar spin swirls. The local magnetic moments of the skyrmion domain could cover the surface of a sphere, giving the topological winding number of skyrmion index. The magnetic skyrmions were theoretically predicted in chiral magnets without inversion symmetry.~\cite{Bogdanov1989} Their existence was later established experimentally in the bulk phases and thin films of non-centrosymmetric B20-type hellimagnets.~\cite{Muhlbauer2009,Yu2010,Yu2011,Adams2012,Seki2012,Tonumura2012,Onoda2009,Butenko2010,Heinze2011,Nagaosa2013}

Skyrmions observed in most of these magnetic systems are of  a two-dimensional  (2D) nature with the constant spin texture along $c$ axis as stabilized on the thin film.  Recently, the three-dimensional (3D) spin density-dependent topological transport phenomena in MnGe indicates  a non-coplanar spin structure;~\cite{Shiomi2013} while the real-space measurement on MnGe demonstrated the stacking of hedgehog and antihedghog spin textures.~\cite{Tanigaki2015} The hedgehog and antihedghog configurations indicate all-out and all-in spin textures with the different sign of the skyrmion index. Therefore, although the other cubic B20 crystals display the 2D skyrmion, MnGe is the unique compound to show the 3D skyrmion, besides high  magnetic ordering temperature~\cite{Takizawa1988,Kanazawa2011,Makarova2012} and small skyrmion size.~\cite{Tanigaki2015,Kanazawa2011}
Understanding of the complicated nature of magnetism in MnGe will be one fundamental challenge of the condensed matter physics. 
Computationally, even in the smallest skyrmion of MnGe among B20 compounds,   the simulation of the 3D skyrmion needs more than 1500 atoms in a supercell built with $6\times 6 \times 6$ primitive cells, which is beyond the current  simulation capability within the {\em ab initio} density functional theory (DFT). Theoretically,  the DFT  could not  capture the magnetic state of some B20 compounds.  For example, the DFT calculation overestimates the magnetic moment of MnSi.~\cite{Jeong04,Choi16} Furthermore, the non-Fermi liquid behavior in MnSi~\cite{Ritz2013} suggests that strong correlation is important to explain the electronic structure of B20 compounds. 
	
In this Article, we explore the origin of 2D and 3D skyrmions and the variation of local moments  in B20 compounds.  We start with the DFT calculations to understand the electronic structure in MnGe. We then  construct an effective low-energy Hamiltonian based on the DFT inputs. Since MnSi and MnGe have the same number of valence states, they can be investigated systematically. We proceed to answer why MnSi and MnGe show the different nature of the skyrmion, within a strong-correlation driven spin-fermion model. We find the origins of 2D and 3D skyrmions and the local moments are controlled by the strength of the coupling between localized and itinerant magnetic moments.

The paper is organized as follows: Section~\ref{methods} introduces
the methodology.
Section~\ref{spin-fermion} introduces and explains the construction of the spin-fermion model. 
In Sec.~\ref{results} we describe computational results.  
Section~\ref{summary} presents summary and concluding remarks.
Additional information is provided in the Appendix.

\section{methods}\label{methods}
We perform the DFT calculations by employing the projector augmented wave method implemented in the Vienna {\it ab initio} simulation package (VASP)~\cite{vasp1,vasp2} and the full-potential linearized augmented plane wave (LAPW) method implemented in the WIEN2k package.~\cite{W2k} We use the generalized gradient approximation (GGA) of Perdew-Becke-Ernzerhof (PBE) for the exchange-correlation functional.~\cite{pbe} The WIEN2k (VASP) package is employed for the primitive (supercell) calculations. We use the experimental lattice parameters and internal atomic positions for 
MnGe.~\cite{Makarova2012}
Through the maximally localized Wannier function method (MLWF)~\cite{Marzari97,Souza01} implemented in Wannier90,~\cite{Mostofi14} the tight-binding Hamiltonian is constructed.  The LAPW results are used as the input for Wannier90.~\cite{Kunes10} The effective model Hamiltonian  is  solved with Tight-Binding Modeling for Materials at Mesoscale (TBM3) package.~\cite{TBM3}

\begin{figure}[tb]
\includegraphics[width=1.0\linewidth]{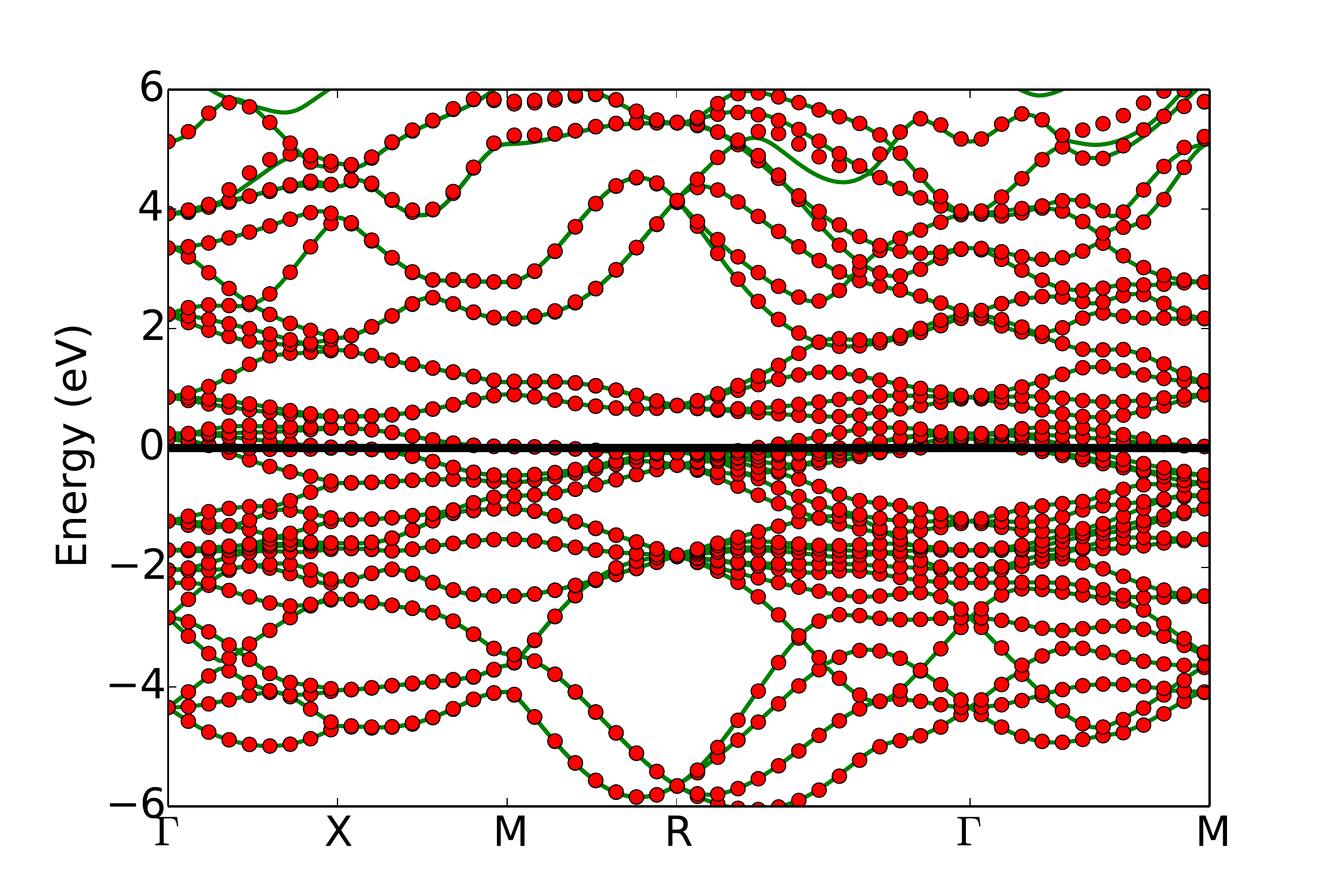}
\caption{(color online)
Overlay of the bandstructure calculated by the DFT eigenvalues (green lines) and  tight-binding model (red dots) in MnGe. The Fermi level is set as 0 eV, which is indicated by the thick black line.
}
\label{fig1}
\end{figure}
 
\section{Spin-fermion model}\label{spin-fermion}
The DFT calculations give a magnetic moment of 2 $\mu_B$  and 1.0 $\mu_B$ per Mn ion in MnGe (4.795 \AA) and MnSi (4.551 \AA), respectively.  Although the DFT gives the result consistent with the experiment~\cite{Makarova2012,Ro$ss$ler2012} for the size of magnetic moment in MnGe, it significantly overestimates the moment size in MnSi, which is found to be only 0.4 $\mu_B$/Mn.~\cite{Ishikawa:1976} To understand this difference, we carried out the DFT calculations for MnGe with the lattice constants of the MnSi crystal~\cite{Jorgensen:1991} and obtained a value of 1.0 $\mu_{B}$/Mn, which is close to that for MnSi. This observation excludes the role of the ligand atom species in causing the drastic moment change. Instead it suggests that the systematic magnetic properties go beyond by DFT and electronic correlation effects must be incorporated into the study of skyrmion properties in these compounds.



It is known that the electronic correlation could produce the dual nature of electrons, showing the coexistence of the localized and delocalized states.  
Although the dual nature is intensively discussed in $f$-electron heavy fermion systems,~\cite{QSi:2004} the concept as represented by the spin-fermion model was also applied to address the quantum critical phenomena in high-temperature cuprates.~\cite{ArAbanov:2003} More recently,  this correlated electron model has also been applied to understand the quantum criticality in Fe-pnictides.~\cite{Dai:2009}
 It is noteworthy that this spin-fermion model has no explicit interaction terms between the local moments after the integration over the incoherent electrons.  This interaction term between incoherent states was reported to play an important role to reproduce the high energy excited state.~\cite{Weicheng:2010} Therefore, this derived spin-fermion model will be appropriate to study the low-energy skyrmion state very close to the magnetic ground state.
Within the correlated electron picture, the electronic excitations encompass an incoherent part far away from the Fermi energy and a coherent part in its vicinity.  The incoherent part corresponds to the lower and upper Hubbard bands in connection with the Mott insulator when the electron on-site repulsion is larger than the Mott localization threshold, and is described in terms of localized magnetic moments; while the coherent part is adiabatically connected to its noninteracting counterpart. It has been shown~\cite{AGeorges:1996}  that this division of the electron spectrum is a successful and convenient way of analyzing the complex behavior of bad metals near the Mott transition.  Here we adopt the same type spin-fermion model to describe  $3d$ electrons in MnGe and MnSi compounds. 

We construct the tight-binding model through the MLWF from the DFT result without the spin-orbit coupling (SOC). Figure~\ref{fig1} shows the MLWFs with Mn $d$ and Ge $p$ well reproduce the DFT band structure between -6 eV and 6 eV. [More information of the electron structure with the DFT calculation is given in Appendices A-C.] Upon a renormalization, this DFT-based tight-binding Hamiltonian represents the coherent part of interacting electrons, which are antiferromagnetically coupled to the localized moments. The system Hamiltonian is written as: 
\begin{eqnarray}
\mathcal{H}&=&\alpha ( H^{hopping}_{TB} +H^{onsite}_{TB}-\mu)+g \sum_{i} \vec{S_i} \cdot \vec{s_i}  \nonumber \\
	&&+h_{B} \sum_{i} S^{z}_{i}\;,
\end{eqnarray}
with 
\begin{equation}
H^{onsite}_{TB} =H^{onsite}_{DFT}	+\sum_{m \in d} \lambda_{d} \vec{l_m} \cdot \vec{s_m} .
\end{equation}
%
%
Here  the renormalization parameter is denoted  by $\alpha$. The variables $\vec{S}$, $\vec{s}$, and $g$, denote the localized and itinerant moments, and the coupling strength between them.
The indices $i$ denote the site, and  $m$, $\vec{l}_m$ and $\vec{s}_m$ indicate $d$-orbital index, its angular momentum quantum number, and its spin quantum number, respectively. The quantities $\mu$ and $h_{B}$ are the chemical potential and  the external magnetic field. $\lambda_{d}$ is the SOC strength and is chosen to fit with the DFT+SOC bandstructure.   Here, $\lambda_{d}$=0.07 eV, and $h_{B}$=0 eV were used.  Since the Hamiltonian has a scaling property with $\alpha$, $g$ and the magnitude of $\vec{S}$ ($|\vec{S}|$), $\alpha=1$ and $|\vec{S}|$=2 were used for convenience.
 
\section{Results}\label{results} 
\begin{figure}[tb]
	\includegraphics[width=1.0\linewidth]{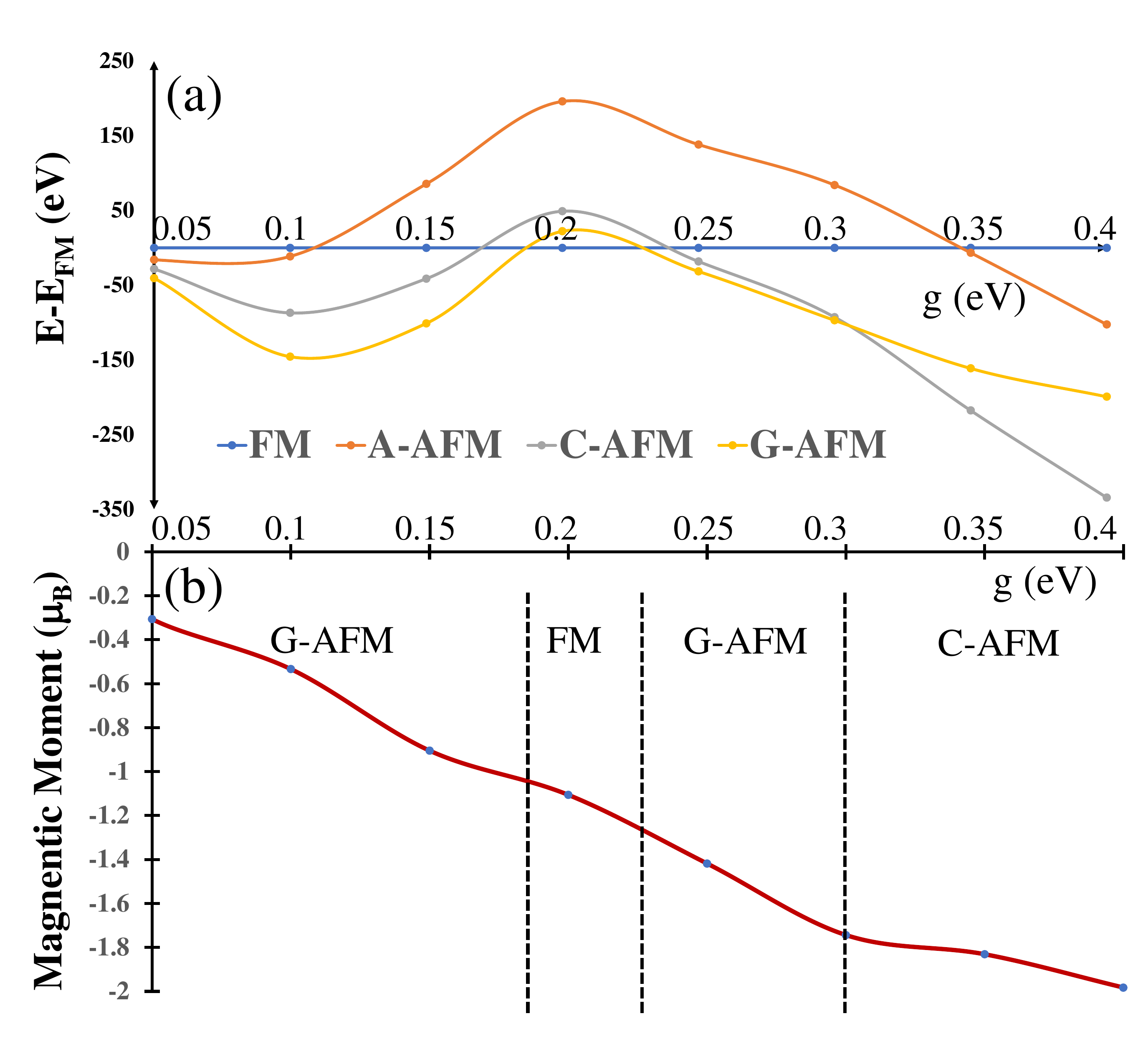}
	\caption{(color online)
	 Total energy (a) for FM, A-AFM, C-AFM, and G-AFM configurations in the $8\times 8\times 2$ supercell and the itinerant magnetic moment (b) as a function of $g$ in the primitive unti cell. In panel (b), the phase transformation with varying $g$ is schematically denoted by dashed lines. 	
	}
	\label{fig2}
\end{figure}

Figure~\ref{fig2}(b) shows  the itinerant magnetic moment as a function of the coupling strength  $g$.  The coupling-driven spin-polarization of the itinerant band  produces an itinerant magnetic moment. For example, the itinerant magnetic moment shows 2.0 (0.3) $\mu_{B}$ at $g$=0.4 (0.05) eV. Because the itinerant magnetic moments is anti-parallel to the classical spin ($|\vec{S}|=2$ $\mu_B$), the total moment at $g$=0.4 (0.05) eV estimates $\sim$ 0 (1.7) $\mu_B$. 
Therefore, the reduced (large) magnetic moment observed in MnSi (MnGe) could correspond to the case of large (small) $g$. This is because $g$ is proportional to the electron hopping, which is enhanced with volume collapse (14.5\% from MnGe to MnSi).  In addition, our results also explain the observation of a significant moment suppression in MnGe under a 6 GPa pressure.~\cite{Ro$ss$ler2012,Deutsch:2014} 


We now examine the total energies for the ferromagnetism (FM), the A-type antiferromagnetism (A-AFM), the C-type AFM (C-AFM), and the G-type AFM (G-AFM ;see Appendix D for the detail of the magnetic configurations) as a function of $g$  in an $8\times8\times2$ supercell.  Since the B20 structure has 4 different transition metal layers stacked along the $c$-axis, the shortest periodicity of $c/4$ along the $c$-axis could be defined.  The possible AFM periodicity along the $c$-axis is the multiple of $c/4$ such as $c/2$, $c$, and $2c$ (see  Fig.~\ref{SFig4} in the Appendix). We found  that $2c$ of the magnetic periodicity along the $c$-axis is suitable to describe the phase transition between C-AFM and G-AFM states (see Fig.~\ref{SFig5} in the Appendix) and  $2c$ was used to make an AFM along the $c$-axis in our study.
The magnetic phase diagram as a function of $g$ between 0.05 eV and 0.4 eV is summarized in Fig.~\ref{fig2}(a). When $g$ is between 0.4 eV and 0.3 eV, the ground state is the C-AFM phase. The crossover between G-AFM and C-AFM phases occurs at $g=$ 0.3 eV. Therefore, G-AFM becomes the ground state at $g<$ 0.3 eV.  The energy of the FM is higher than other configurations at large ($>0.35$~eV) and small ($<0.1$~eV) $g$.  For $g$ between 0.1 eV and 0.35 eV, the A-AFM shows the highest state among them.  As $g$ approaches 0.05 eV, the difference in the total energies of four magnetic states  becomes very tiny, because the magnitude of itinerant moments becomes negligible at small $g$.  

The phase transition between C-AFM and G-AFM could provide the useful insight into the 2D and 3D skyrmions. The center and boundary of the skyrmion show the antiferromagnetic relation. If the distance between the center and border becomes as short as possible, the skyrmion would be comparable to an antiferromagnetism. 
Since the G-AFM state has the alternating spin directions layer by layer, it might be associated with the 3D skyrmion at the extremely small size. 
Therefore, as shown in  Fig.~\ref{fig2}(a) and 
Fig.~\ref{fig4}(a), one expects that 3D (2D) skyrmion can emerge  when $g$  is smaller (larger) than 0.3 eV.
It is noteworthy to mention that the three orthogonal helices could also describe a three-dimensional spin texture.  As the G-AFM and C-AFM could be defined in terms of orthogonal helices,  the low-energy spin-wave theory of skyrmion crystal based on the orthogonal helices~\cite{Zhang:2016}  might shed light on the relation between the conventional AFM and the skyrmion crystal.


\begin{figure*}[tb]
	\includegraphics[width=1.0\linewidth]{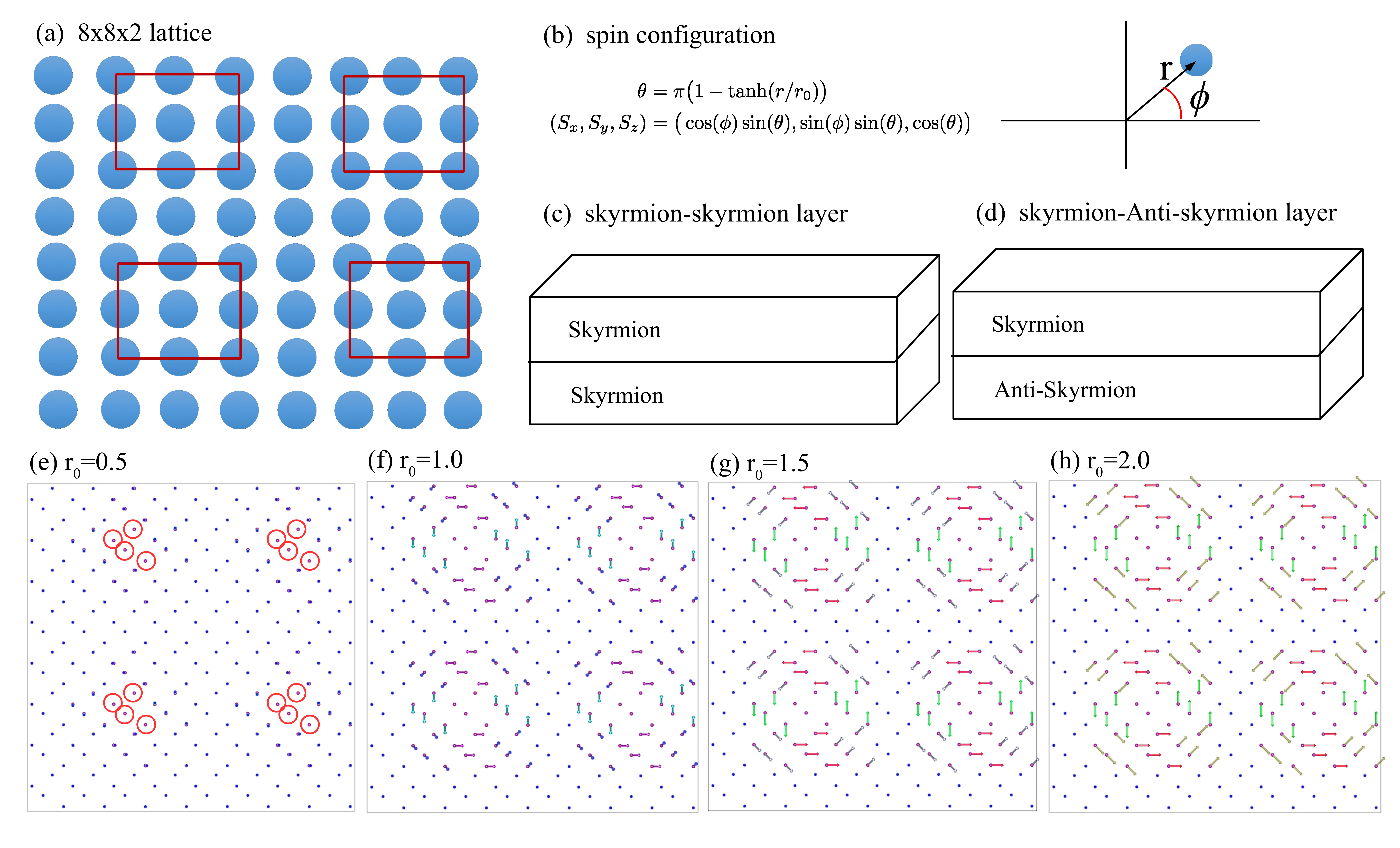}
	\caption{(color online) 
	 Schematic spin structure of the skyrmion. (a) The skyrmion spin structure was placed within the red rectangles in the $8\times8\times2$ supercell. (b) Skyrmion spin formula. (c) Skyrmion-skyrmion lattice. (d) skyrmion-anti-skyrmion lattice. (e)-(h) the different size of skyrmion in the $8\times 8\times2$ supercell. The neighbor 4 red circles represent the top view of Mn atoms in the primitive unit cell of MnGe at the center of a given red rectangle in (a). The skyrmion means the positive skyrmion index, and the anti-skyrmion means the negative skyrmion index.
	}
	\label{fig3}
\end{figure*}

Based on the above insight, we now investigate 2D and 3D skyrmion properties in real space. A skyrmion lattice is constructed in  the $8\times8\times2$ supercell shown in Fig.~\ref{fig3}(a). The single skyrmion is manipulated inside each rectangle with $(\cos(\phi) \sin(\theta),\ \sin(\phi) \sin (\theta),\ \cos (\theta))$. The distance $r$ and the azimuthal angle $\phi$ are computed with respect to the center of each rectangle as depicted in Fig.~\ref{fig3}(b). $\theta$ is a function of $r$ for a given parameter $r_0$, characterizing the size of the skyrmion texture. Every classical spin at the center ($r=0$) of the rectangle points downward. Outside each rectangle, all classical spins are aligned upward to satisfy the boundary condition of skyrmion. For example, for $r_0$=0.5, only the  spin at the center points downward while others becomes almost upward. The different sizable skyrmions are shown in Fig~ \ref{fig3}(e)-(h) as a function of $r_0$. The skyrmion index is the summation of the solid angle ($\Theta$) over the spin texture.  We used the following formula to calculate the solid angle:
\begin{equation}
\Theta_i = 2\arctan\left[\frac{\mathbf{S}_{i} \cdot (\mathbf{S}_{j} \times \mathbf{ S}_{k})} {1+\mathbf{S}_i\cdot\mathbf{S}_j +\mathbf{S}_j\cdot\mathbf{S}_k +\mathbf{S}_k\cdot\mathbf{S}_i} \right],
\label{sk-solid}
\end{equation}
subtended by three neighboring spins, $\mathbf{S}_i$ in the 2D plane. The skyrmion index is given by $\sum_i\Theta_i/4\pi$ in each Mn layer. The skyrmion index of each layer needs to be $+4$ or $-4$ due to the four skyrmion in each layer.  Practically, the skyrmion indices are $-4.0, \;-3.93,\;-3.76,\;-3.56$,  for  $r_0=0.5,\;1.0,\;1.5,\;2.0$, respectively. The non-integral topological index suggests a finite size effect that the rectangles as marked in Fig.~\ref{fig3}(a) cannot fully accommodate a larger-sized skyrmion. However, because the tiny skyrmion would be stabilized in MnGe, this finite size effect does not occur to MnGe.  
The 2D skyrmion could be easily placed using the equivalent skyrmion formula as a function of a layer. The sign of the skyrmion index was determined by the sign of the numerator of Eq.~(\ref{sk-solid}). Due to the multiplication of three spins in the numerator, the whole sign change of the classical spins layer by layer drives the sign change of the skyrmion index along  the $c$ axis. 
Here the positive skyrmion index means a skyrmion and the negative one implies an anti-skyrmion. Although the observed 3D skyrmion has the alternating stack of the hedgehog (all-out) and antihegehog (all-in) textures, we build the  3D skyrmion structures by stacking 2D skyrmion or anti-skyrmion planes along $c$-axis.
There is another way to generate an anit-skyrmion by the inverted vorticity.~\cite{Nayak:2017,Hoffmann:2017} To show an analogy between the 3D skyrmion and the G-AFM spin texture, we inverted spin directions to generate an anti-skrymion.

\begin{figure}[tb]
	\includegraphics[width=1.0\linewidth]{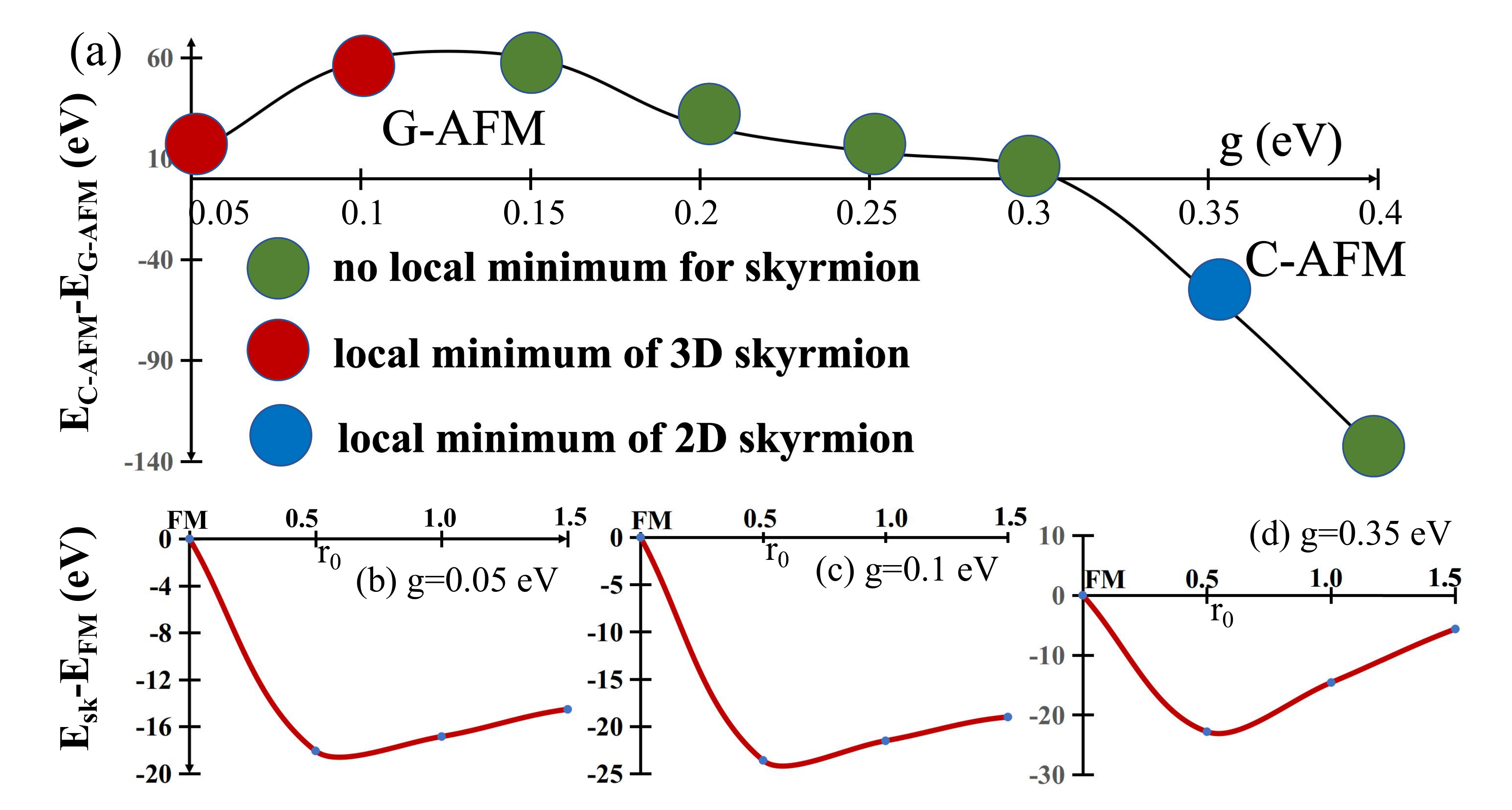}
	\caption{(color online)
  		Total energy difference between G-AFM and C-AFM states  as a function of the coupling strength $g$ (a). 
		The total energy of the 3D (b-c) and 2D skyrmion (d), relative to the FM state, as function of $r_0$. 
		The color of circles represents whether a local minimum of 3D (red) or 2D (blue) exist or not (green).
		Here the values of $g=0.05$ (b),  0.1 (c)  and  0.35 (d) eV were used in the calculations. The local minimum were obtained for 3D (b)-(c) and 2D (d) skyrmions. 
			}
	\label{fig4}
\end{figure}

We searched the local minimum of the total energy of the 2D or 3D skyrmion as a function of $g$ and $r_0$  in reference to the FM energy. 
Figure~\ref{fig4}(b)-(d) present the relative total energy of 2D or 3D  skyrmions as a function of $r_0$ at $g$= 0.05 eV, 0.1 eV, and 0.35 eV. We find a stabilized 2D skyrmion at $g$=0.35 eV and stabilized 3D skyrmions at $g$ = 0.05 eV and 0.1 eV. Figure~\ref{fig4} (b) and (c) demonstrate that the 3D skyrmion with $r_0$=0.5 is stabilized at $g < 0.3 \;\text{eV}$. Figure~\ref{fig4}(d) shows the 2D skyrmion with $r_0$=0.5 is stabilized at  $g> 0.3\; \text{eV}$.  In other parameters, we could not find a stabilized skyrmion in comparison to the FM energy. Our model shows the very tiny skyrmion in both the stabilized 2D and 3D skyrmions.  Since  our tight-binding model  is constructed from the DFT inputs of MnGe, large-sized skyrmions as observed  in MnSi could not be stabilized with the current parameters.

It is noteworthy that, while the DFT calculations always obtain that the FM state has a lower  energy than any skyrmion configuration in MnGe, the spin-fermion model indeed predicts several important results: (1) the stabilized skyrmion state in comparison to the FM state, (2) the phase transition from 2D to 3D skyrmions with reduced $g$, and (3) the skyrmion lattice in the 8$\times$8$\times$2 supercell. 
To test the stability of the skyrmion state,  the local minimal of the 2D ($g =0.35\; \text{eV}$ and $r_0=0.5$) and 3D ($g =0.1 \;\text{eV}$ and $r_0=0.5$)  skyrmions are iterated by Langevin-Landau-Gilbert (LLG) spin dynamics, implemented in TBM3 package:~\cite{TBM3}
\begin{align}
\frac{d\vec{S}_i}{dt}=\vec{S}_i \times \vec{F}_i+\eta (\vec{S}_i \times \vec{F}_i) \times \vec{S}_i \;.
\label{LLG}
\end{align}
where $\vec{F}_i$ is the effective field from the Hellmann-Feynman
theorem.  In Eq.~(\ref{LLG}), $\eta$ is a positive value for the damping term and we set $dt = 0.1$ and 0.02 to update the local spin orientation for $g$=0.05 eV and 0.35 eV, respectively. The time-dependent evolutions of the local minimums would confirm these skyrmions are indeed stable.  The same skyrmion index was maintained until the magnetic state was converged. With the slight modification of the magnetic structure, the initially-imposed skyrmions were stabilized within the criterion of 0.000001 $\mu_{B}$ for the difference of each magnetic moment.

\section{Summary}\label{summary}
We have demonstrated that the spin-fermion model with the antiferromagnetic coupling between the itinerant and localized electrons can capture the magnetic properties of MnGe and MnSi at the same time. The model is based on the tight-binding model from the DFT result in MnGe.  At large values of coupling strength $g$, the compensation of the localized and itinerant moment leads to a reduced moment state  (MnSi) and gives rise to a 2D skyrmion. At small values of $g$, the reduction of itinerant moment gives a large moment state (MnGe) and  a 3D skyrmion.  The compensation-induced small magnetic moment state at large g could be comparable to the compensated magnetic moment of the majority and minority spin parts at the large onsite Coulomb interaction in MnSi.~\cite{Collyer:2008,Shanavas:2016}
 We  have found the itinerant moment controlled by $g$ plays an important role in determining whether  a 2D or 3D skyrmion should be stabilized.  Our spin-fermion model has given a consistent picture on the understanding of the 2D and 3D skyrmions in B20  compounds.
Since the three-dimensional spin configurations are shown to be associated with the magnetic state of MnGe, we might need to reconsider multi-magnetic periodicity (mutli-$Q$ state) beyond the single magnetic periodicity that occurs in the helical ground state.~\cite{Koretusne:2015}

\acknowledgments
We acknowledge useful discussions with Shi-Zeng Lin  and Zhoushen Huang. This work was supported by U.S. DOE  Contract No. DE-AC52-06NA25396 through the LANL LDRD Program  (H.C. $\&$ Y.-Y.T.) and U.S. DOE BES Program under LANL E3B5 (J.-X.Z.). The work was supported in part by the Center for Integrated Nanotechnologies, a U.S. DOE BES user facility, 
in partnership with the LANL Institutional Computing
Program for computational resources.

\appendix
\section{Bulk DFT calculation}
\begin{figure}[tb]
	\includegraphics[width=1.0\linewidth]{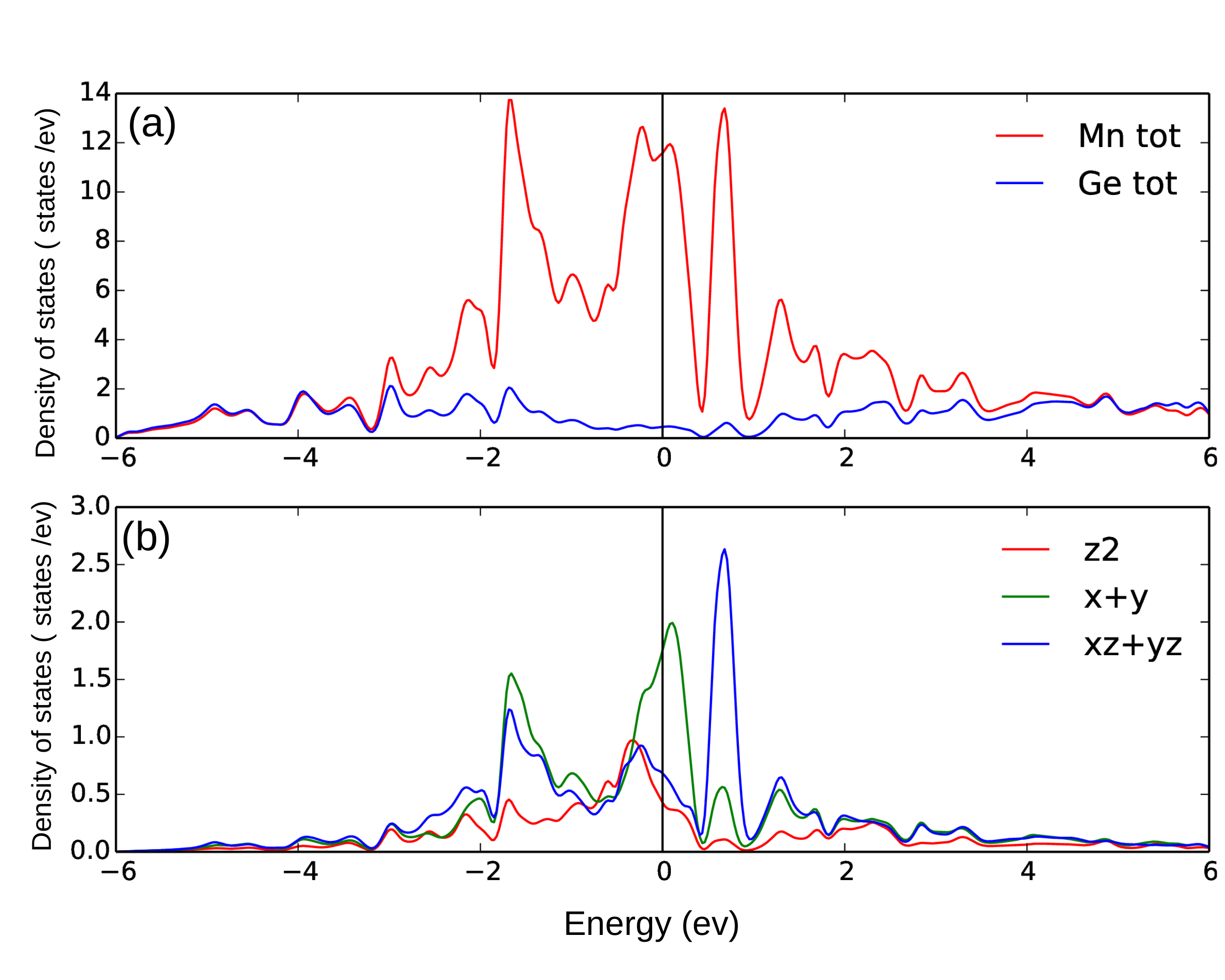}
	\caption{(color online)
		Total density of states for Mn (red) and Ge  (blue) atoms (a), and partial density of states for $3d$ electrons in Mn atoms (b).  The local symmetry of Mn atoms induces $3d$ orbital states to be split into $d_{z^2}$ (red), $d_{x+y}$ (green), and $d_{xy+yz}$ (blue). 0 eV is set as the Fermi level.
	}
	\label{SFig1}
\end{figure}

We performed the DFT calculation in the non-magnetic MnGe. Figure~\ref{SFig1} shows the density of states (DOS) in non-magnetic MnGe. With the three-fold rotation symmetry along  the (1,1,1) direction, $3d$ orbital states In Mn ions could be split into $d_{zx}$+$d_{yz}$,$d_{x^2 - y^2}$+$d_{xy}$ and $d_{z^2}$, whose DOS are presented in Fig.~\ref{SFig1}(b). All $d$  states are mainly distributed between [-2 eV, 2 eV]. Also, there is the strong hybridization between Mn $3d$ and Ge $4p$ states, as shown in Fig.~\ref{SFig1}(a). The  Ge-$4p$ partial DOS intensity at $E_F$ is too small compared with that of Mn $3d$. Therefore, Mn $3d$ states should  have a major role for the magnetism.

\section{Bandstructures with and without spin-orbit coupling}
\begin{figure}[tb]
	\includegraphics[width=1.0\linewidth]{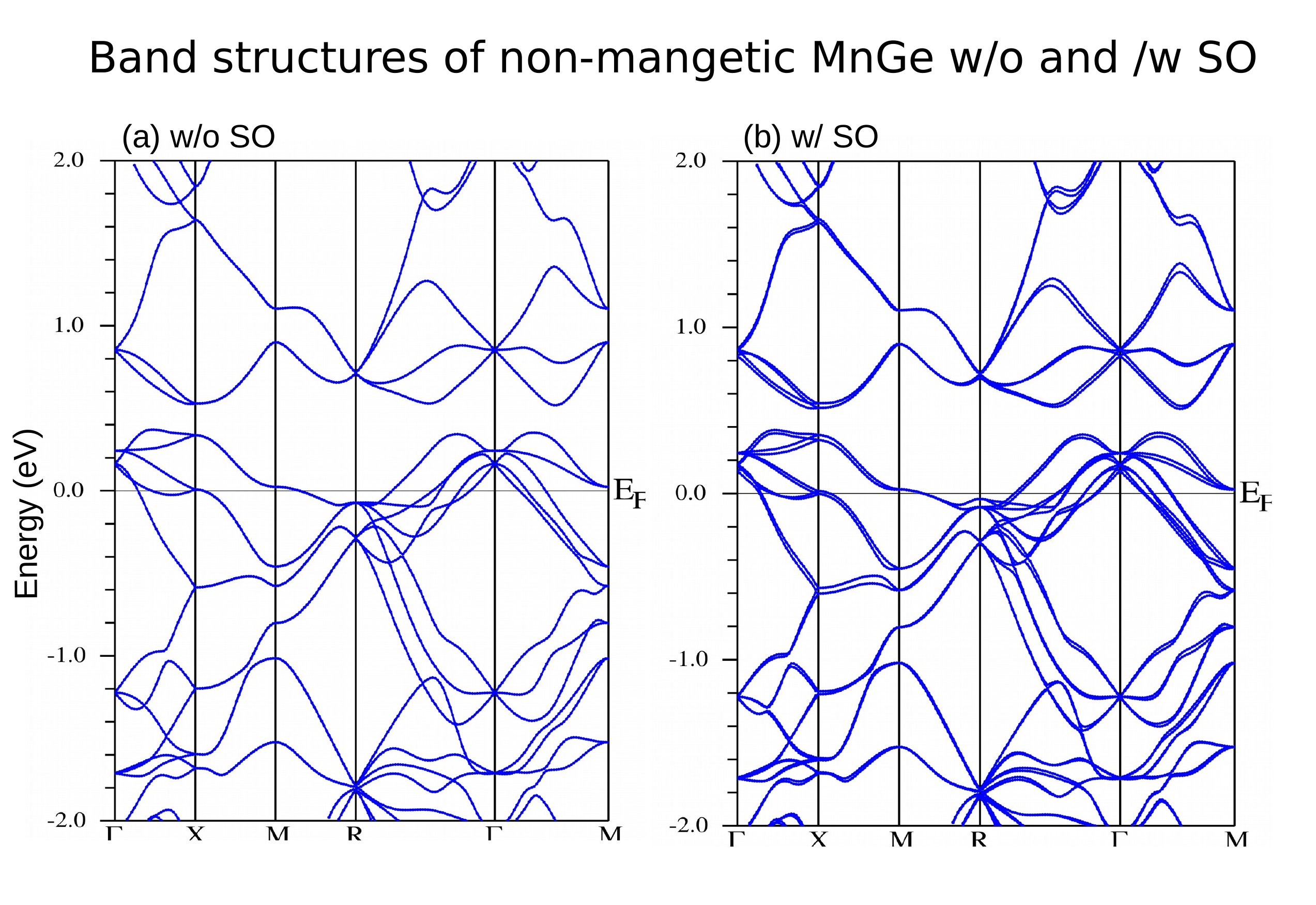}
	\caption{(color online)
		Bandstructue of non-magnetic MnGe (a) without and (b) with spin-orbit coupling. 
	}
	\label{SFig2}
\end{figure}
We investigated the spin-orbit coupling (SOC) effect on MnGe. Figure~\ref{SFig2} shows the band structure with and without SOC. The energy splitting due to SOC is about 1 meV.  This strength of the SO is well fitted with  $\lambda_d$=0.07 eV in $\lambda_{d} \vec{l}_m \cdot \vec{s}_m$.

\section{Total energies as function of skyrmion size}
\begin{figure}[tb]
	\includegraphics[width=1.0\linewidth]{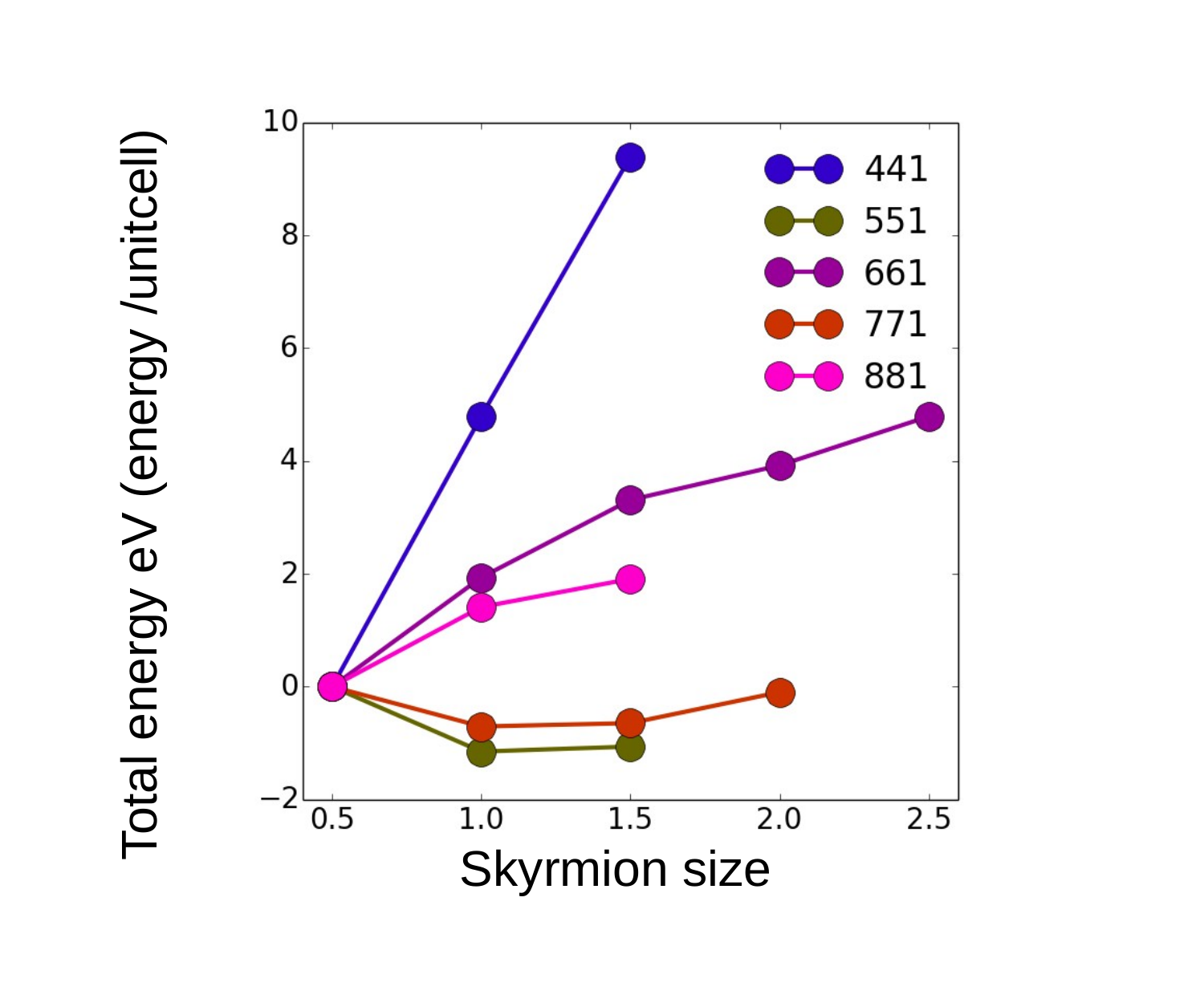}
	\caption{(color online)
		Relative total energy of a skyrmion  as a function  of  the  initial  skyrmion  size $r_0$ in  the  $n\times n\times 1$  supercell ($n=5$,$7$). Here we take the energy of the skyrmion phase with $r_0$=0.5 as an energy reference.
	}
	\label{SFig3}
\end{figure}
 We performed the full-relativistic non-collinear DFT calculation of the skyrmion spin texture in MnGe. The spin-orbit coupling was taken into account in the calculations. Using the skyrmion definition in Fig.~3 (b), we put the 2D skyrmion texture in $n \times n \times 1 $ supercells ($n$=integer). Figure~\ref{SFig3} shows the total energy as a function of skyrmion size in different supercells. Only $5\times5\times1$ and $7\times7\times1$ supercell could have a stabilized skyrmion with $r_0$ =1. The calculated size of skyrmion is smaller than $r_0$=2 in MnSi.~\cite{Choi16}
Both DFT calculations on MnGe and MnSi show the energy of skyrmion with respect to the ferromagnetic state would be larger by about one to two meV/f.u. (Specifically, 1.67 meV/f.u. in MnGe and 0.84 meV/f.u. in MnSi). This energy scale in MnSi was discussed to show a comparable estimate to the experimental observation.~\cite{Choi16}

\section{Magnetic configurations}
\begin{figure}[tb]
	\includegraphics[width=1.0\linewidth]{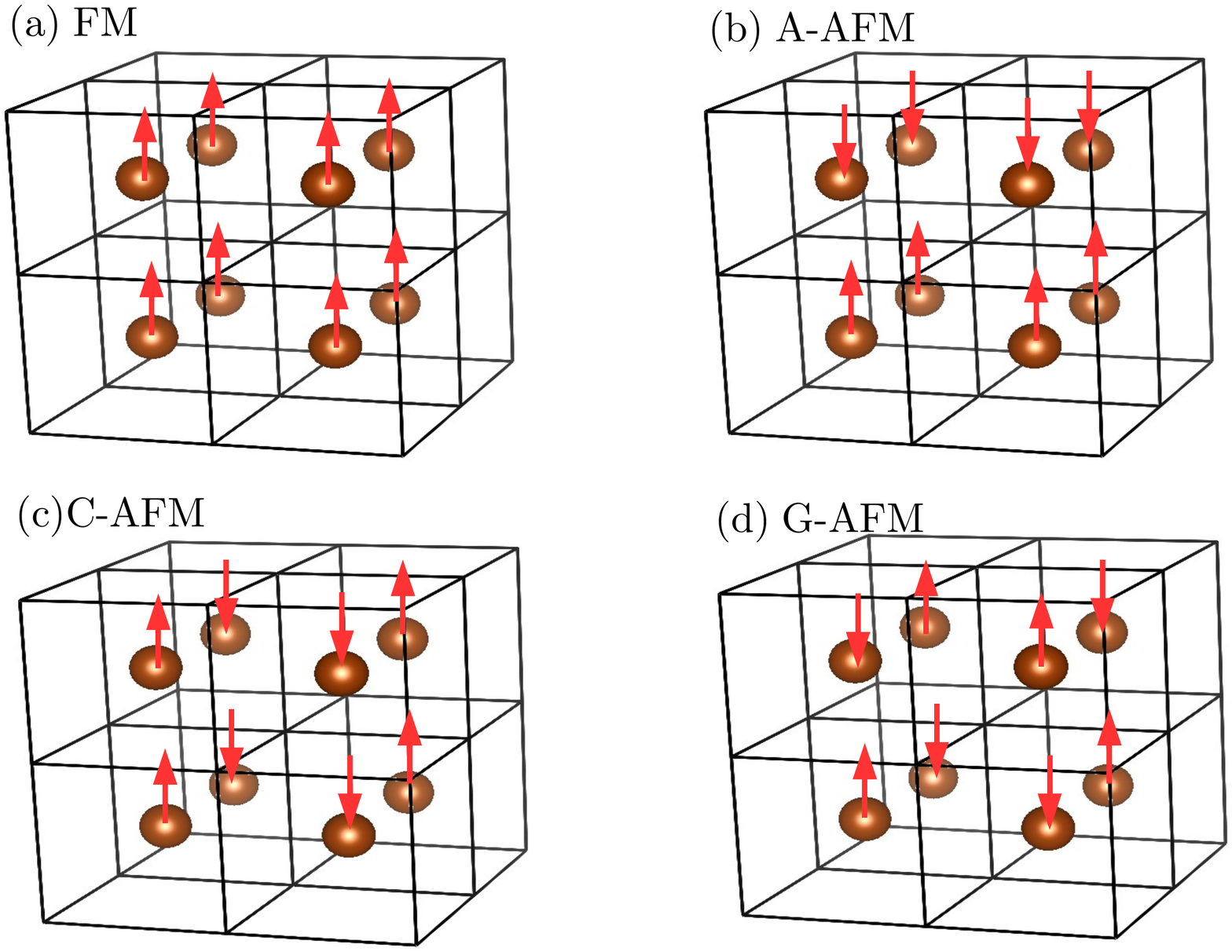}
	\caption{(color online)
		Schematic magnetic structures. (a) the ferromagnetism (FM) (b) A-type antiferromagnetism (A-AFM) (C) C-type antiferromagnetism (C-AFM) (d) G-type antiferromagnetism (G-AFM).The all four Mn atoms have the identical moments in the unitcell of MnSi.
	}
	\label{Magntism}
\end{figure}

FM, A-AFM, C-AFM, and G-AFM configurations have the conventional magnetic configurations with q=(0,0,0), (0,0,pi), (pi,pi,0), (pi,pi,pi), respectively. The typical structures are given in Fig.~\ref{Magntism}.

\section{Periodic boundary condition}
\begin{figure}[tb]
	\includegraphics[width=1.0\linewidth]{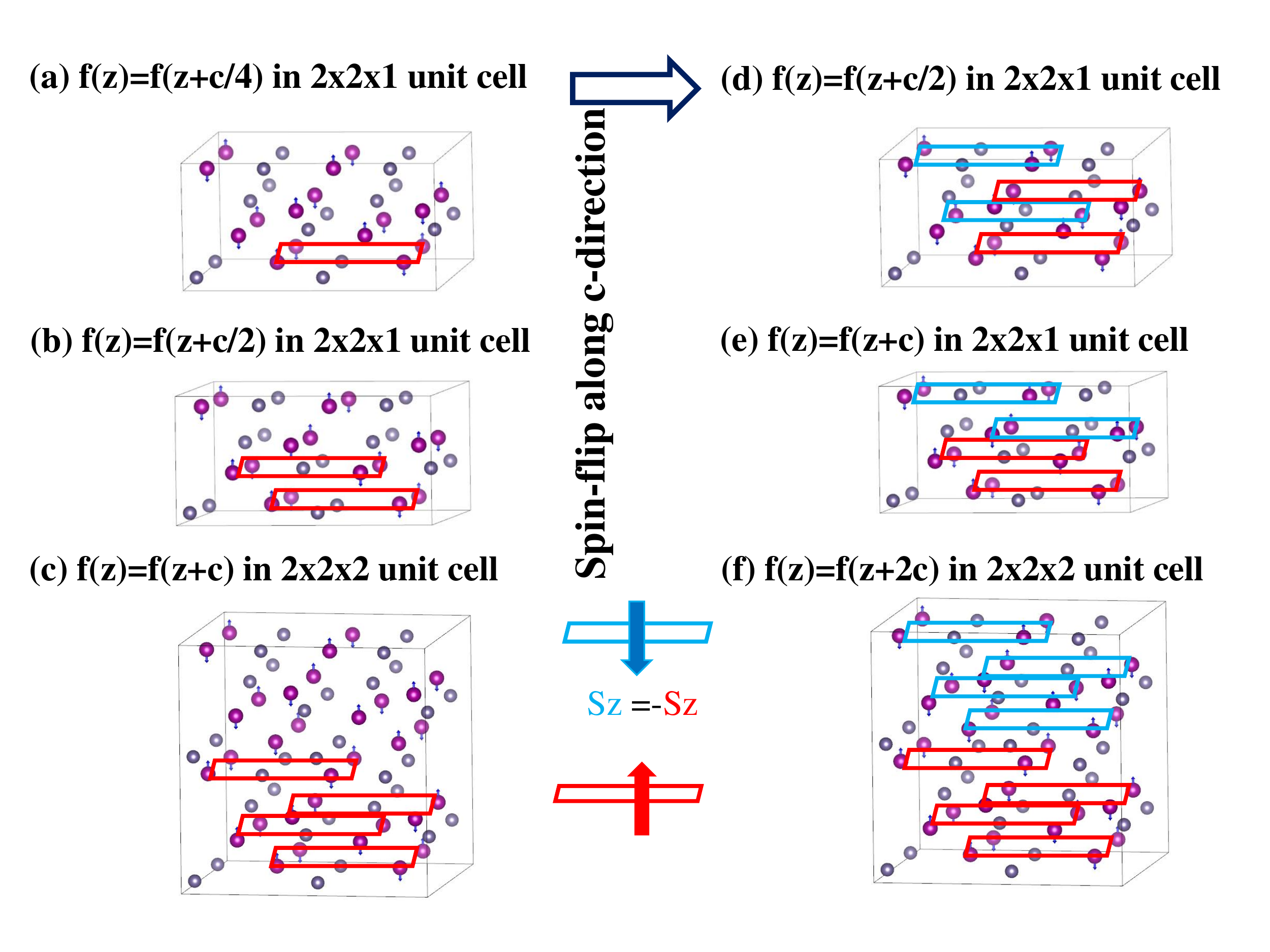}
	\caption{(color online)
		Schematic spin structure of G-type antiferromagnetism with different periodic boundary conditions. The magnetic unit cell becomes doubled along the $c$-axis with the spin-flip along the $c$-axis.  Panels (a)-(c) show the different periodic boundary conditions along the $c$-axis. After the spin-flip along the $c$-axis, Panels (d)-(f) show the doubled periodic boundary conditions. 
	}
	\label{SFig4}
\end{figure}
There are several choices to make a G-AFM in the B20 structure.  The primitive unit cell of MnGe has four Mn-Ge layers stacked along the $c$ axis. We could define the shortest periodicity of $c/4$ along the $c$-axis. The AFM periodicity along the $c$-axis could be the multiple of $c/4$ such as $c/2$, $c$, and $2c$.   Here we used the spin arrangement shown in Fig.~\ref{SFig4}(f) to achieve 3D skyrmions. The reason of our choice will be presented below.

\begin{figure}[tb]
	\includegraphics[width=1.0\linewidth]{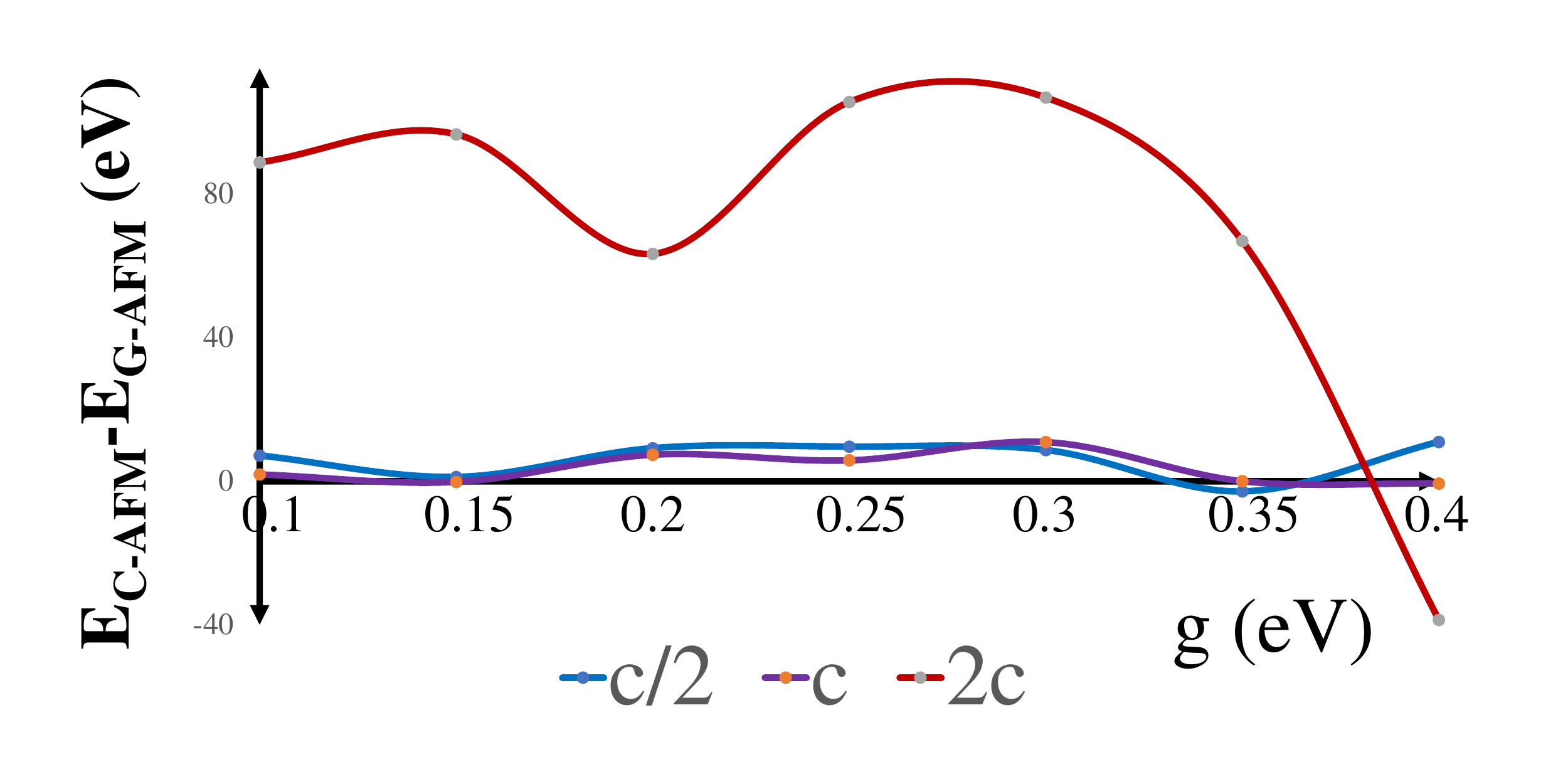}
	\caption{(color online)
		The difference between G-AFM and C-AFM as a function of $g$ for different magnetic periodicity ($c/2$, $c$, $2c$) along the $c$-axis within the $1\times1\times1$ momentum mesh. 
	}
	\label{SFig5}
\end{figure}

We performed the total energy calculation of G-AFM and C-AFM states in the $8\times8\times2$ supercell with the  $1\times1\times1$ momentum mesh. Figure~\ref{SFig5} shows that the periodicity of $2c$ along the $c$-axis produces the clear phase transition between G-AFM and C-AFM states. Therefore, a $2c$ length was used for the antiferromagnetic periodicity along the $c$-axis throughout this work. We used a $4\times4\times2$ momentum mesh to simulate the skyrmion properties in the spin-fermion model in this work due to the convergence requirement.

\end{document}